\documentclass[structabtsract,usenatbib]{an}
\usepackage{graphicx}
\usepackage{natbib}
\usepackage{color}
\def\teff{$T_{\rm eff}$}
\def\kms{km\,s$^{-1}$}
\def\degs{$^{\circ}$}
\def\msun{M$_{\odot}$}         \def\rsun{R$_{\odot}$}
\def\lsun{L$_{\odot}$}

\def\fracp#1/#2{\leavevmode\kern.05em\raise.6ex\hbox{\the\scriptfont0
    #1}\kern-.15em/\kern-.15em\lower.3ex\hbox{\the\scriptfont0 #2}}

\begin{document}

\titlerunning{HD 69479}

\title{Composite Spectra, Paper XXIII: HD 69479/80,}
\subtitle{a 90-day binary with a cool-giant primary}

\author{R.E.M. Griffin \inst{1}
\and R.F. Griffin \inst{2} }
\institute{Herzberg Astronomy \& Astrophysics Research Centre, NRC, Victoria, 
BC, Canada \\
   \email{Elizabeth.Griffin@nrc-cnrc.gc.ca}
\and Institute of Astronomy, The Observatories, Cambridge,  UK \\
   \email{rfg@ast.cam.ac.uk}  }

\date{Received  }

\abstract{HD~69479/80 is a composite-spectrum binary whose components are a
  late-G giant and an early-A dwarf.  The orbit has a period of only 91 days
  (which seems short for a system containing a cool giant with a radius of
  $\sim$13 solar radii), and a very small, but probably significantly non-zero,
  eccentricity. We separated the component spectra by a procedure of spectral
  subtraction, using a standard single giant spectrum as a template, and found
  the closest match to the spectrum of the cool component to be that of 15~Cyg
  (G8\,III). We measured the radial velocity of the secondary component from
  each uncovered spectrum, solved the SB2 orbit for the system, and derived a
  mass ratio $m_1$\,/\,$m_2$ of 1.318. Fitting synthetic spectra to the spectra
  of the secondary component indicated a \teff~of 9250\,K, {\it log}\,g = 3.75,
  and a rotational velocity of $\sim$90\,\kms.  We determined the difference in
  absolute magnitude, $\delta$V, between the component stars to be 1.07 mag,
  the late-type component being the brighter; we could thence calculate radii
  and luminosities for both components, plot their H--R diagram positions, and
  fit evolutionary tracks.  The best-fitting tracks indicated masses of
  2.9\,\msun~for the giant and 2.2\,\msun~for the dwarf, which was fully in
  keeping with the mass ratio given by the SB2 orbit. The track for the dwarf
  star confirms that this component has begun to evolve away from the
  ZAMS. Fitting the corresponding isochrone to those H--R diagram positions
  indicated a log\,(age) of the system of approximately 8.60 Gyr since the cool
  star evolved from the ZAMS, which is a little younger than the ages deduced
  for many cool giants.  We also detected the $\lambda$\,6707-\AA~lithium line
  in the spectrum of the giant component, thus adding to the evidence that it
  is near the start of its primary ascent of the red-giant branch.}


\maketitle

\section{Introduction and History}

HD 69479/80 is a 6.5-mag star in the constellation Hydra, about
10\degs\,following Procyon.  The cool component of the system is dominant in
the red and the hot component in the violet, so both spectra are seen in the
near-UV and blue regions.  The compossite nature of the object was first
recognized almost 100 years ago by \citet{Cannon24}, who assigned it two
numbers in the {\it Henry Draper Catalogue}, HD~69479 (type G0) and 69480 (A2).
It has a $V$ magnitude of 6.53 \citep{Corben71}, just marginally fainter than
the formal 6.50 mag limit for the {\it Bright Star Catalogue}
\citep{Hoffleit82}. Efforts at assigning spectral types to the component stars
have generally agreed that the secondary is an early A-type dwarf, but were
somewhat less unified as regards the giant component (see
Table~\ref{tab:types}).  Even quite recently some catalogues have treated
HD~69479/80 as a single star; the {\it Hipparcos} catalogue, for example,
assigns a single number to the object.

\begin{table}
\caption{Previous classifications of HD\,69479/80}
\begin{tabular}[]{lccl} 
\noalign{\smallskip}\hline\noalign{\smallskip}
Reference & Giant & Dwarf~ & \llap{No}tes \\
\noalign{\smallskip}\hline\noalign{\smallskip}
Cannon \& Pickering (19\rlap{24)} & G0 & A2 & 1 \\
Adams et al. (1935) & \multicolumn{2}{c}{dF8} & 2 \\
Markowicz (1969) & G5\,III & A2\,V: \\
Olsen (1979) & cF6 & Ap & 3  \\
Eggen (1986) & K0\,III & A2\,V & 4 \\
Houk \& Swift (1999) & K0\.III--IV & A2\,V & 5 \\
Ginestet et al. (1997) & G7\,III$^{-}$& & 6 \\ 
Ginestet et al. (2002) & G7\,III$^{-}$& A2\,(IV:) & 7a \\
{\it loc.~cit.} & G9:\,III--IV & A2\,V & 7b \\
\noalign{\smallskip}\hline\noalign{\smallskip}
\end{tabular}
\label{tab:types}
\vskip6pt
{\it Notes:} \\
1. HD classification \\
2. Also called ``{\it Composite}''.  $M_V$ = +3.6 \\
3. From Str\"omgren indices \\
4. ($B-V$) = 0.61 mag; {\it cf.} ($B-V$) = 0.63 from {\it Hipparcos} \\
5. Prism spectra (Michigan re-classification programme) \\
6. Near-IR classification (prism) spectra \\ 
7a. $\Delta\,M_v$ = 0.65 mag \\
7b. Model, using the parallax and $(B-V)$ from {\it Hipparcos} \\
\vspace{-20pt}
\end{table}

The spectroscopic model prepared by \citet{Ginestet02} was based on a distance
modulus of 7.47 mag derived from the {\it Hipparcos} parallax of 3.22 $\pm$ 0.6
mas. It yielded an absolute magnitude for the system of --0.94 (which could be
considered a little high, given the absence of evidence of raised luminosity;
see Table~\ref{tab:parms}), a combined ($B-V$) of 0.55 mag, individual absolute
magnitudes of 0.0 mag for the giant and +0.8 mag for the dwarf, and individual
($B-V$) values of 0.91 mag (giant) and 0.05 mag (dwarf).  Comparison with the
{\it Hipparcos} ($B-V$) of 0.63 mag indicated a reddening $E_{(B-V)}$ of 0.08
mag, or an extinction $A_V$ of 0.28 mag, which is in keeping with the distance
of 311 $\pm$ 64\,pc indicated by the {\it Hipparcos} parallax.

\begin{figure*}
\protect\vspace{-30pt}
\hskip-40pt
\hspace{-50pt}
\includegraphics[width=1.5\linewidth]{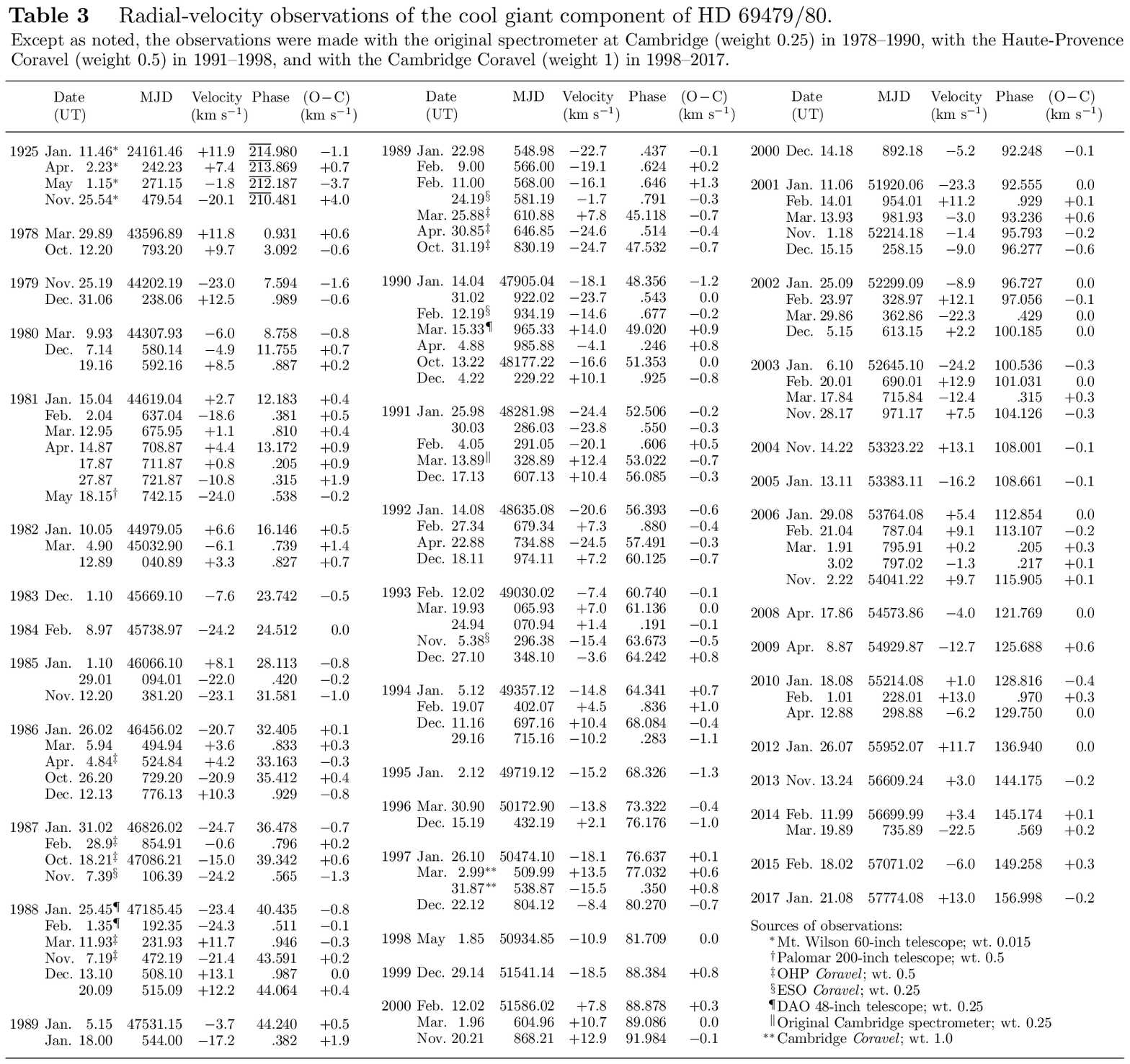}
\label{tab:RV-t3}
\end{figure*}
\setcounter{figure}{0}

This present analysis is based on a parallax of 4.368 $\pm$ 0.07 mas measured
by {\it Gaia} \citep{Gaia DR2-18}, which indicates a distance modulus of 6.80
mag; the corresponding observed absolute magnitude $M_V$ of --0.27 mag can be
considered more characteristic of a system containing a modest late-type giant
and an early-type dwarf (see Section 5.1 and Table~\ref{tab:parms}). Since the
star is therefore now thought to be 36\% nearer and correspondingly less
luminous, the stellar parameters need to be revised downwards (see Section
5.2). Appeal to the semi-empirical three-dimensional model of Galactic
interstellar extinction \citep{Arenou92} suggests that the amount of extinction
at the position and distance (229 pc) of HD~69479/80 is small: $A_V$ = 0.09 mag, or a reddening $E_{(B-V)}$ = 0.03 mag.

\section{Radial Velocities and Orbit of the Cool Star}

On the basis of unpublished Mt.~Wilson radial velocities, \cite{Hynek38} had
suggested that the radial velocity of HD~69479/80 is variable. In a major and
very public-spirited effort, \cite{Abt70,Abt73} listed the radial velocities,
measured from about 23,000 Mt.~Wilson plates, that had been published
previously only as mean values for the stars concerned.  In the case of
HD~69479/80 there were four velocities, all obtained in the year 1925 and
having a range as great as 37~\kms, thus offering strong confirmation of its
binary nature. Other publications are scarce; \cite{Duflot95} listed a mean of
0~\kms~for the system, derived from four individual plates, but did not give
dates.

It was because of the reported composite nature of the spectrum that
HD~69479/80 was placed on the Cambridge radial-velocity (RV) programme in 1978.
In the following year the velocity was seen to have changed by 32\,\kms, and
from then on it has been measured routinely. \citet{Griffin90} published
orbital elements for the late-type component of the binary, though without
listing the data or undertaking any discussion. Although both spectra are
visible in the mid-blue, the RV spectrometers used here were optimised for
observing cool-star spectra, and thus did not detect useable signals from the
hot star.  The RVs referred to here are therefore those of the cool component
alone, and yielded an SB1 orbit (see Table~\ref{tab:SB1}).

119 RV observations with full metadata have now been measured for the cool
component.  They are set out in Table~3: 115 of our own observations and the
four from Mt.~Wilson published by Abt (1970). Of our own observations, most
(38) of the early ones were measured with the original photoelectric
spectrometer \citep{Griffin67} on the Cambridge 36-inch reflector; 31 others
were obtained on a guest-investigator basis by RFG with the Haute-Provence
(OHP) {\it Coravel} \citep{Baranne79}, four with the analogous instrument at
ESO, three with the RV spectrometer \citep{Fletcher82} on the 48-inch Dominion
Astrophysical Observatory (DAO) telescope, and one with the Palomar RV
spectrometer \citep{Griffin74}.  Nearly all (38) of the observations obtained
since 1997 have been made at Cambridge with an RV spectrometer based on the
{\it Coravel} design.

The RVs measured at OHP and ESO (and reduced by Geneva's own software) have
been adjusted by adding 0.8\,\kms, an amount determined empirically to
represent the systematic difference between the Geneva and the Cambridge
scales. (That step is justified, since the OHP measurements had been subject to
an undisclosed colour-dependent zero-point correction applied at source.)  This
adjustment has been made to all the relevant observations prior to their entry
in Table~3.

To bring the weighted residuals from the SB1 solution into tolerable equality,
the new {\it Coravel} observations have been weighted by 1.0, those from OHP
and Palomar by 0.5, those from the original spectrometer, ESO and the DAO by
0.25, and the old Mt.~Wilson observations by 0.015.  The resulting orbital
elements are listed in Table~\ref{tab:SB1}, where they are compared to the
elements published by \citet{Griffin90}.  While the new values of the elements
are little changed, their precisions are substantially tighter by virtue of the
methodology of the new Cambridge {\it Coravel}.  We have added to Table~3 the
phases and residuals of the observations as given by the orbit solution in
Table~\ref{tab:SB1}.

\begin{table}[h!]
\centering
\caption{SB1 orbit solution for the giant component
\label{tab:SB1}}
\begin{tabular}[]{lcll}
& & Griffin (1990) & This paper \\
\noalign{\smallskip}\hline\noalign{\smallskip}
$P$ (days) &  &  90.836 $\pm$ 0.006 & 90.8406  $\pm$ 0.0014 \\
$T$ (MJD)  &  &  46147.12 $\pm$ 0.12 & 51143 $\pm$ 0.9 \\
$\gamma$ (\kms)  &  &  --5.52 $\pm$0.10 & $-$5.65 $\pm$ 0.05 \\
$K$ (\kms) &  &  18.77 $\pm$ 0.14 & 18.72 $\pm$ 0.07 \\
$e$        &  & 0 & 0.0060 $\pm$ 0.0035 \\
$\omega$   &  & --  & 359 $\pm$ 36 \\
$a $\, sin\,$i$ (Gm)  &  & 23.44 $\pm$ 0.18 & 23.38 $\pm$ 0.09 \\
$f(m)$ (\msun)    &  & 0.0624 $\pm$ 0.0014 & 0.0619 $\pm$ 0.0007 \\
\multicolumn{3}{l}{{\it R.m.s.} residual (wt.~1)} 0.7~~~~~~~~~~~ &  0.37 \kms \\
\end{tabular}
\end{table}

\section{Spectroscopy of HD~69479/80}

\begin{figure*}
\vskip-60pt
\hskip-120pt
\includegraphics[height=16in]{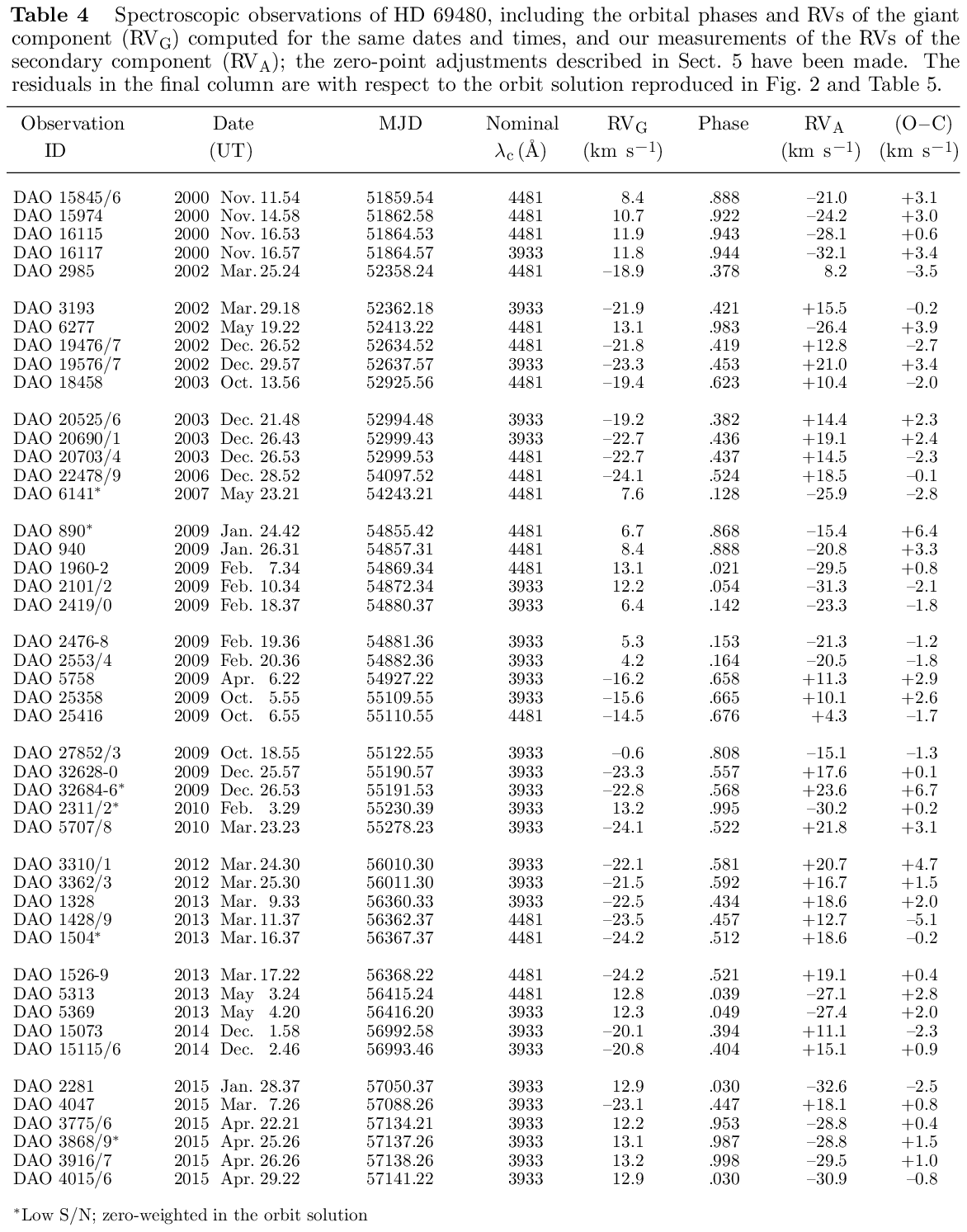}
\caption{Table 4~~Spectroscopic observations of HD\,69480, including the
orbital phases and computed RVs of the giant component for the corresponding 
dates and times, along with the measured RVs of the secondary component.  The 
residuals in the final column are with respect to the orbit solution in 
Table~5 and Fig.~2. }
\label{tab:log}
\end{figure*}
\setcounter{figure}{0}
\setcounter{table}{4}

An exploratory photographic spectrogram of the star was obtained at
10\,\AA\,mm$^{-1}$ at the coud\'e focus of the Mt.~Wilson 100-inch telescope in
1981.  It at once revealed the composite nature of the object: a somewhat
rotationally broadened Ca~{\sc ii} $K$ line typical of an early-A star, amid a
crowd of narrow lines characteristic of a cool giant. Beginning in 2000, the
system was monitored at the coud\'e focus of the 1.2-m DAO telescope and
96-inch spectrograph, with a resolving power of ~$\sim$45,000. The spectrograph
employs a Richardson image slicer \citep{Richardson68} and a {\sc sit}e-4 CCD,
and delivers images of spectra as {\sc fits} files.  Between 2000 and 2016 94
useable exposures were obtained in the spectral regions of interest here; about
half were made in sequential pairs and were co-added in order to improve the
S/N ratio of the output, thus reducing the actual numbers of useable spectra
to 50 (28 near the $K$ line, 4 at H$\delta$ and 18 in the blue near
4481\,\AA). Each spanned about 145 \AA.  Table~4 lists the observations that
were used in this analysis, together with the orbital phases and RVs of the
primary on the corresponding dates as given by the single-lined orbit solution.

\subsection{Uncovering the spectrum of the hot component}

\begin{figure*}[!t]
\centering
\includegraphics[height=15.0cm,angle=-90]{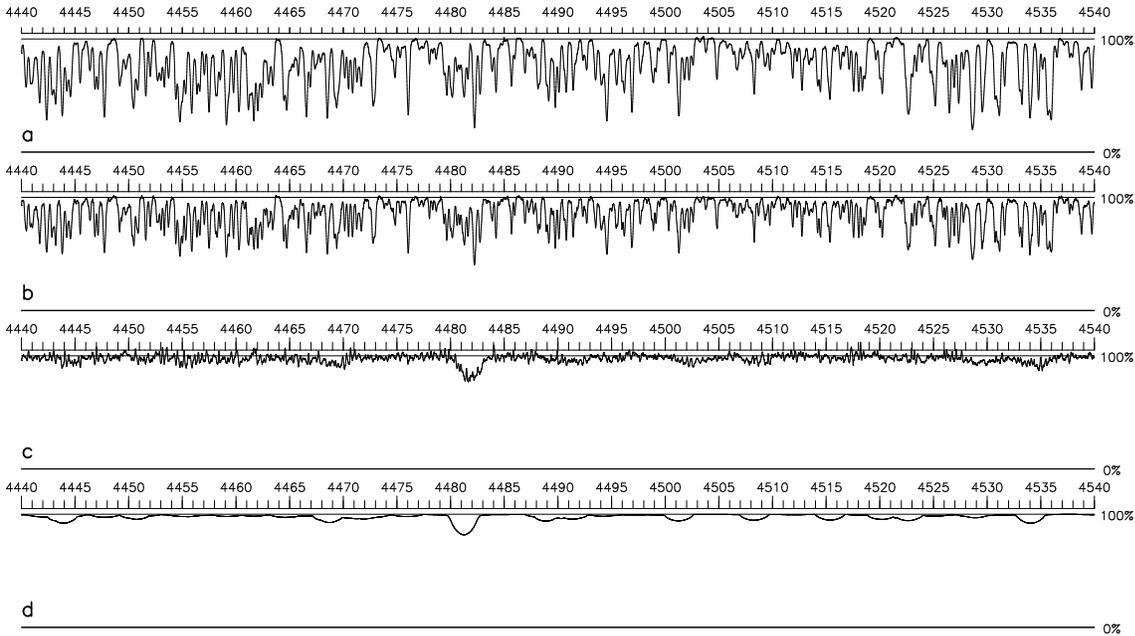}
\caption{Uncovering the spectrum of HD~69480. When the spectrum of the 
G8\,III standard 15~Cyg (panel {\bf a}) is subtracted optimally from one of the
composite spectra (panel {\bf b}), the spectrum of the A2-type secondary 
(HD~69480) is revealed (panel {\bf c}).  It is compared in panel {\bf d} with a
synthetic spectrum calculated for \teff~= 9250\,K and log\,{\it g} = 3.75, and
blurred to mimic a rotational velocity of 90\,\kms.}
\label{fig:sub}
\end{figure*}

To separate the superimposed spectra of these binaries we adopt a technique of
point-by-point subtraction.  By appealing to trial and error methods we
determine the best match to the spectrum of the primary from among a library of
cool-giant spectra.  A good match is indicated when there are very few
(preferably no) residual artefacts (`spikes') in the secondary's spectrum
caused by mis-matches in strength or profile (width) between the giant's
spectrum lines and those of the surrogate.  Note that it is highly important to
align the spectra of the giant component and of the standard giant precisely in
wavelength to avoid the creation of unwanted artefacts in the residue arising
from small discrepancies between wavelength scales. To that end, we routinely
extract the raw (observed) spectra as intensities per pixel, and apply the
grating equation as appropriate for the DAO spectrograph to derive a wavelength
solution.  All our spectra are thus in the rest-frame of the standard (or cool
giant) star.
     
In the present case (as illustrated in Fig.~\ref{fig:sub}), the spectrum of the
primary star (HD~69479) was matched closely by that of the cool giant standard
15~Cyg, classified by \citet{Keenan89} as G8\,III.  A small disparity in
line-widths, in the sense that the lines of 15~Cyg are a little broader than
those of the HD~69479, was removed by blurring the spectra of the latter by a
rotational velocity of 4~\kms.  The residue of the subtraction -- the spectrum
of the hot component, HD~69480 -- was compared with synthetic models; it was
fitted best by one calculated for \teff~= 9250\,K, log\,$g$ = 3.75 and solar
abundances, derived from an {\sc Atlas9} model and Gray's {\sc Spectrum}
package.\footnote{\tt http://www.physics.appstate.edu/spectrum/spectrum.html}

\section{RVs of the Secondary Star, and a Double-Lined Orbit}

We measured the RVs of the secondary star by cross-correlating each extracted
spectrum with the synthetic spectrum described above and broadened to match the
line-widths of the secondary star. Since the composite spectra were extracted
in the rest-frame of the giant component (as just explained), the RVs
measured for the secondary needed to be corrected for the RV of the giant on
the date of each observation (column 5 of Table 4), as determined from the SB1
orbit (Table~\ref{tab:SB1}).

The observations of the H$\delta$ region were helpful for determining the
\teff~of the uncovered secondary star, but H$\delta$ was the only measureable
feature in that region and we could not determine reliable RV measurements from
it, so we have not included those observations in Table~4.

\begin{figure}[!b]
\vskip-20pt
\hskip-30pt
\includegraphics[height=8.3cm]{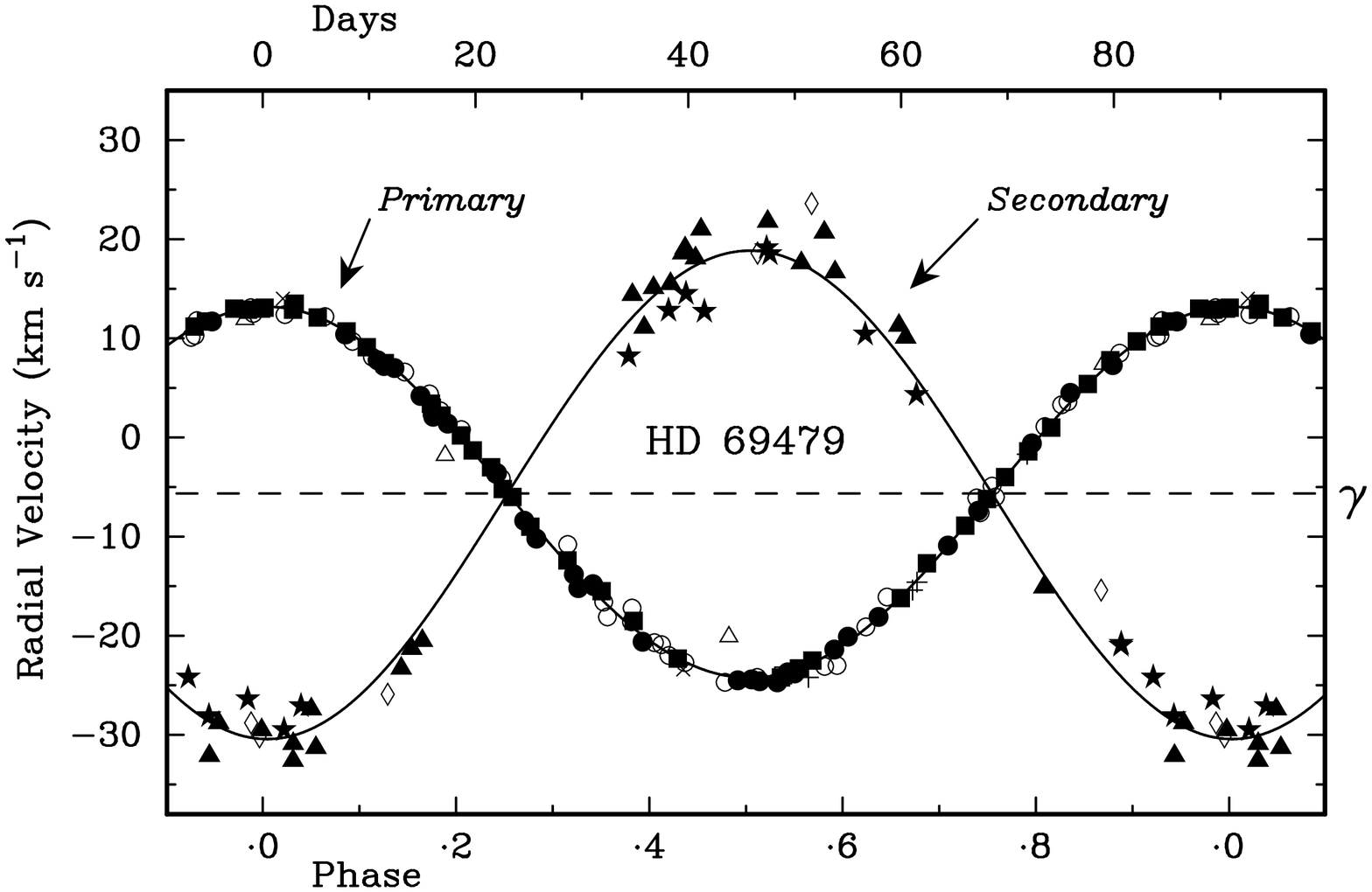}
\vskip-30pt
\caption{Double-lined orbit of HD~69479/80.  The symbols denote (for the giant)
  the origin of each measurement, and (for the dwarf) the wavelength region
  involved.  The sources of the RVs and their respective weightings are as 
follows: 
}
\includegraphics[width=16.0cm]{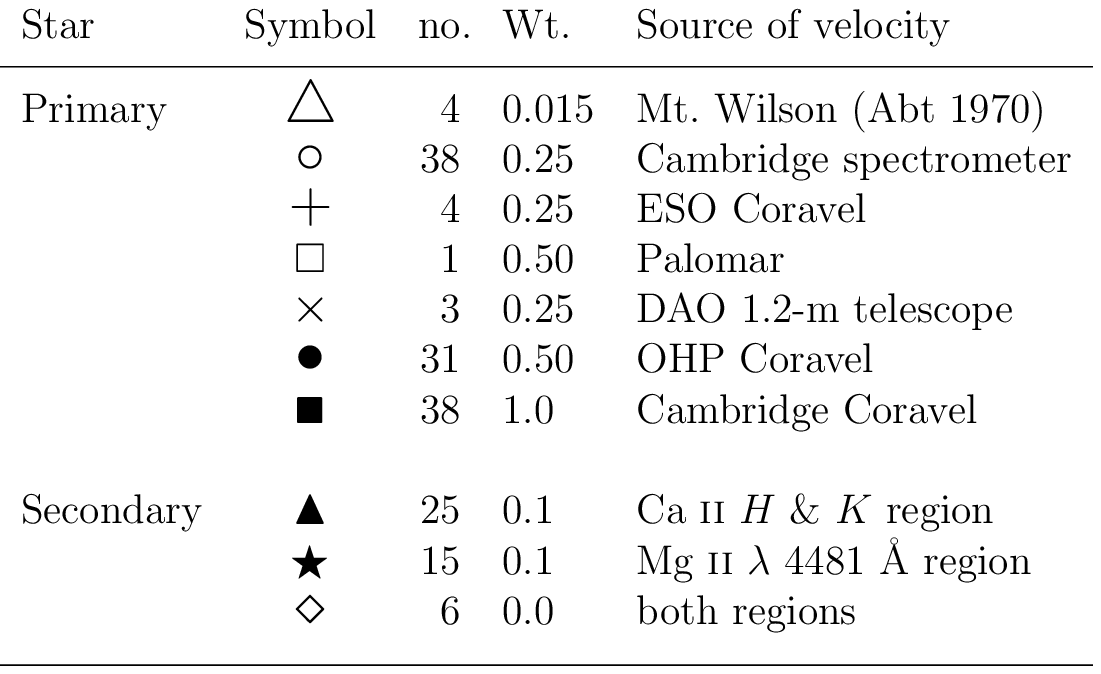}
\vskip-500pt
\label{fig:SB2}
\end{figure}

RV measurements of the spectrum of the early-A star (HD~69480) are difficult to
make very accurately.  Not only do early-A stars have rather few spectral lines
even in the near-UV, but this A-type star is also rotating by about 90\,\kms, causing
its lines to be broad and shallow and making the weaker ones too indistinct to
measure well -- or even to detect with much certainty.  The only prominent
lines in the near-UV region that we observed were two Balmer lines ($H\epsilon$
at $\lambda$\,3970\,\AA~and H8 at $\lambda$\,3889\,\AA) and the Ca\,{\sc ii} $K$
and $H$ lines at $\lambda$\,3933~and 3968\,\AA.  The Balmer lines are intrinsically wide and
therefore not well suited for RV measurements; moreover, H8 is right at the
short-wavelength end of our spectra, where the stellar flux gradients are
changing rapidly owing (largely) to atmospheric transparency and the CCD
response, while $H\epsilon$ is closely blended with the Ca\,{\sc ii} $H$ line
and is consequently additionally complicated by whatever asymmetries affect the $K$ line.

We experimented with cross-correlating the entire length of each near-UV
spectrum of HD~69480, but concluded that the most consistent set of
measurements were those that involved just the $K$ line.  We carried out
similar trials with various sub-sets of lines in the blue region; however,
since the signal in the observed spectrum had been reduced to about only 35\%
of the original level by our subtracting away the contribution of the giant,
the noise level in the secondary spectra was inevitably raised, and the most
reliable cross-correlation signals were obtained when the wavelength region was
limited to the Mg~{\sc ii} $\lambda$\,4481\,\AA~line.  The RVs thus derived
(and recorded in Table~4) could consequently be affected by individual
systematic errors, which we attempted to minimise by obtaining a substantial
number of observations, many in sequential pairs with the intention of
increasing the S/N levels of the spectra by co-adding them. Unfortunately,
because the star is a winter object and at a declination of only +4\degs,
favourable observing conditions for it in the winter climate of Victoria are
not very plentiful.

At its distance of 229 pc (Section~1), HD~69479/80 could show weak interstellar
($IS$) absorption, though we found no clear evidence of it. $IS$ features can
be detected best in early-type stars, whether or not rotating.  In
composite-spectrum systems, $IS$ $K$ lines have readily been detected and
measured in the uncovered spectra of a hot secondary if its spectral type is
earlier than A3; examples are HD~4615 \citep[][Fig.~3]{Griffin99}; HR~233 and
36~Tau \citep[][Figs.~5 \& 7]{Mason97}.  In the case of HD~69479/80 the
secondary's $K$ line is strong enough, and therefore broad enough, to mask a
weak $IS$ line; furthermore (as already explained) the S/N ratios in our
near-UV spectra were rather low, and we did not identify any features that
could definitely be attributable to $IS$ absorption and not to random noise
spikes.  As mentioned in Section~1, the amount of $IS$ absorption calculated
for this system from a semi-empirical model is small, and our deduction via a
quite different route (see Table~\ref{tab:parms}) confirms that finding.

We combined our measured RVs of the secondary star with those of the primary as
recorded in Table~3, to derive the SB2 orbit for the system.  It is illustrated
in Fig.~\ref{fig:SB2}; the elements of the orbit are listed in
Table~\ref{tab:SB2}.  The caption of Fig.~\ref{fig:SB2} links the various
symbols to the sources of RVs for the primary component, or -- in the case of
the secondary component -- to the respective wavelength region observed.  As is
our custom in these derivations of SB2 orbits, we applied a global weight of
0.02 to all the secondary velocities.  Observations of the secondary that did
not achieve a satisfactory level of exposure were zero
weighted; they are plotted as open diamonds.

The secondary's velocities exhibited a small systematic offset from the SB1
solution for the giant component, and a universal correction of --3.0 \kms~was
applied to all of the secondary's RVs so as to minimize the residuals compared
to those of the giant star.  Those corrections have been applied to the RVs of
the secondary prior to their entry in column 7 of Table~4.  The residuals
listed in the final column of that Table are with respect to our SB2 solution.

\begin{table}
\centering
\caption[]{Orbital elements for HD~69479/80$^*$
\label{tab:SB2}}
\begin{tabular}[]{ll}
\vspace{-0.15in} \\
\noalign{\smallskip}\hline\noalign{\smallskip}

\vspace{-0.1in} \\
$P$ (days)        & =~  90.841 $\pm$ 0.001 \\
$T$ (MJD)         & =~  51052 $\pm$ 9 \\
$\gamma$ (\kms)   & =~  --5.64 $\pm$ 0.05 \\
$K_1$ (\kms)      & =~  18.72 $\pm$ 0.07 \\
$K_2$ (\kms)      & =~  24.6  $\pm$ 0.5  \\
$q$        & =~  1.316 $\pm$ 0.027 \\
$e$        & =~  0.006 $\pm$ 0.003 \\
$\omega$ (degrees)        & =~  359 $\pm$ 35 \\
\vspace{-0.1in} \\
$a_1$\, sin\,$i$ (Gm) & =~  23.38 $\pm$ 0.08 \\
$a_2$\, sin\,$i$ (Gm) & =~  30.8 $\pm$ 0.6 \\
$f(M_1)$ (\msun)  & =~  0.062 $\pm$ 0.001 \\
$f(M_2)$ (\msun)  & =~  0.141 $\pm$ 0.008\\
$M_1$\,sin$^3$\,$i$ (\msun) & =~  0.437 $\pm$ 0.021 \\
$M_2$\,sin$^3$\,$i$ (\msun) & =~  0.332 $\pm$ 0.008 \\
\vspace{-0.1in} \\
\multicolumn{2}{l}{R.m.s. residual (wt.~1) (\kms) = 0.36} \\

\noalign{\smallskip}\hline\noalign{\smallskip}
\multicolumn{2}{l}{$^*$Subscripts 1 and 2 denote the primary and the} \\
\multicolumn{2}{l}{secondary star, respectively} \\
\vspace{-0.1in} \\
\end{tabular}
\end{table}

\section{Parameters of the Component Stars}

\subsection{Photometric model}

As a by-product, the subtraction procedure furnishes the ratio of the fluxes
from the component stars at specified wavelength intervals.  By selecting
50-\AA~bands centred on steps 25\,\AA~wide we can make use of the photometry of
a range of bright stars of various luminosities published by
\citet{Willstrop65}.  The goal is to find a pair of stars that resemble the
binary components sufficiently well as to give a constant
wavelength-independent fit to the ratios of their fluxes.  Since Willstrop's
magnitudes were normalized in $V$, the fraction of the published values
required to match those of the binary yields the value of $\Delta V$ between
the two component stars.  In this way we derived $\Delta V$ = 1.07~mag for
HD~69479/80.

\begin{table}[b]
\centering
  \caption{Published {\it T$_{\rm eff}$} for 15~Cyg}
\begin{tabular}[]{ll} 

Value & Reference \\
\noalign{\smallskip}\hline\noalign{\smallskip}

4953 $\pm$ 30   & Takeda et al.~(2008) \\
5004 $\pm$ 32   & Stock et al.~(2018) \\
5042 $\pm$ 26   & Wu et al.~(2011) \\
5050 $\pm$ 80   & Hekker \& Melendez (2007) \\
5085 $\pm$ 85   & Luck \& Heiter (2007) \\
\end{tabular}
\label{tab:teff}
\end{table}

\begin{table*}
\hskip50pt
\begin{minipage}{16.0cm}
\caption[]{Physical parameters of the component stars of HD~69479/80
\label{tab:parms}}
\begin{tabular}{lrcrrrrrrcr}
\vspace{-0.1in} \\
\noalign{\smallskip}\hline\noalign{\smallskip}
Component Star & $M_V^{~*}$~ & ($B-V$) & \teff & BC\,~ & $M_{\rm bol}$
  & $R$~~& log\,$L$ & $M$~ \\
  & (mag\rlap{)} & (mag\rlap{)} & (K)  & (mag)
  & (mag\rlap{)}&({\rsun}\rlap{)} & (\lsun) & (\msun) & \\
\vspace{-0.1in} \\
\noalign{\smallskip}\hline\noalign{\smallskip}
\vspace{-0.1in} \\
Primary (G8\,III) & 0.07 & 0.95 & 5050 & $-$0.29  & --0.22
  & 12.9  & 1.98 & 2.9 \\
 HD 69479 & \hskip7pt \llap{$\pm$}0.04 & 
  & $\pm$70
  & \llap{$\pm$}0.03 & \llap{$\pm$}0.04 & \llap{$\pm$}0.4 & \llap{$\pm$}0.02
  & \llap{$\pm$}0.1 \\
Secondary A2\,IV & 1.14 & 0.03 & 9250 &  --0.11 & 1.03
  & 2.2  & 1.48 & 2.2 \\
HD 69480 & \hskip7pt \llap{$\pm$}0.04 &
& $\pm$200
  & \llap{$\pm$}0.03  & \llap{$\pm$}0.04 & \llap{$\pm$}0.1 & \llap{$\pm$}0.02
  & \llap{$\pm$}0.1 \\ \\

System (\it modelled) & --0.27   & 0.61  &\\
\hskip32pt(\it observed) & --0.27\rlap{$^*$} & 0.63  & \\
                      & \hskip7pt \llap{$\pm$}0.04 \\ \\

\multicolumn{10}{l}{$^*$Based on the \citet{Gaia DR2-18}
distance modulus of 6.80 $\pm$ 0.04 mag.} \\
\vspace{-0.1in} \\

\noalign{\smallskip}\hline\noalign{\smallskip}
\end{tabular}
\end{minipage}
\end{table*}

The parallax of HD~69479/80, measured by \citet{Gaia DR2-18} as 4.368 $\pm$
0.070 mas, corresponds to a distance of 229 $\pm$ 4 pc and a distance modulus
of 6.80 $\pm$ 0.04 mag.  If $m_V$ = 6.53 mag (Section~1), the absolute magnitude
$M_V$ of the system is --0.27 mag.  The values of $M_V$ for the component stars
and their individual and combined (modelled) ($B-V$) colours are listed in
Table~\ref{tab:parms}. The results for ($B-V$) also support the thesis that
there is only a very small amount of $IS$ absorption affecting this system.

It is more than a little disconcerting that the {\it Gaia} parallax is so
different from the {\it Hipparcos} one, which is listed as 3.22 $\pm$ 0.60 mas
\citep{vL2007}; the difference is almost twice the formal error of the {\it
  Hipparcos} value. One possible cause of erroneous parallax measurements in
this system is a confusion between the orbital motion and the motion of the
photocentre.  That could be significant in this case because the orbital period
is only 0.4 day shorter than a quarter of a year. The likelihood of interplay
between the two orbits will depend on the cadence of the observations;
measurements made only at the optimal times (i.e., at 6-monthly intervals) will
have occurred at the same phases of the binary and will not have been affected
by relative movements of the photocentre.  Unfortunately, information regarding
the dates of the observations was not readily available.

Comparing the modelled value of ($B-V$) given in Table~\ref{tab:parms} with the
{\it Hipparcos} one of 0.63 mag indicates $E_{(B-V)}$ = 0.04 mag (i.e., $A_v$ =
0.13).  Both that and the value given by Arenou et al.'s semi-empirical model
suggest that interstellar absorption is not very significant for this system.

\subsection{\teff~and luminosities}

In these analyses we customarily adopt for the giant component the value of
\teff~that has been derived for the star selected as the surrogate giant for
the subtraction process, in this case 15~Cyg. The five independent published
values listed in Table~\ref{tab:teff} point to a mean of 5050\,K and an
estimated precision of $\pm$ 70 K, so that was adopted for the giant component.
A \teff~of 9250\,K for the secondary had already been determined by comparing
its spectra with synthetic ones, as outlined in Section 3.1.

From the \teff~and $M_V$ for both stars we calculated their luminosities,
radii and $M_{bol}$.  The results are recorded in Table~\ref{tab:parms}. 

\subsection{Evolution and age}

\begin{figure}[!b]
\centering
\includegraphics[height=7cm,angle=-90]{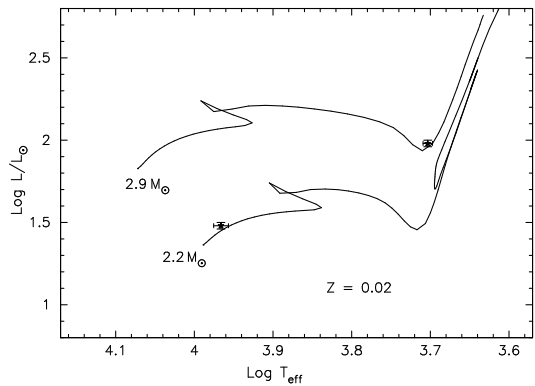}
\caption{Evolutionary tracks, created by \citet{Pols98} and made available in the public domain, have been fitted to the H--R positions of the binary's
  components stars.  Solar abundances were assumed in both cases.  The
  best-fitting curves were those calculated for 2.9\,\msun (giant) and
  2.2\,\msun (dwarf), i.e., a ratio of 1.318, which is acceptably close to the
  value for $q$ of 1.316 given by the orbit solution.  }

\label{fig:evol}
\end{figure}

\begin{figure}[!b]
\centering
\includegraphics[height=7cm,angle=-90]{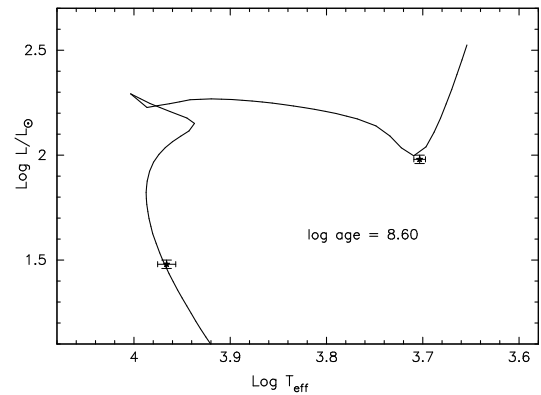}

\caption{An isochrone from \citet{Pols98}, derived from Fig.~\ref{fig:evol} and
  fitted to the H--R positions of the component stars, indicates an interval of 
8.60 Gyr since the more massive star left the main sequence.  }

\label{fig:iso}
\end{figure}

Evolutionary tracks, created by \citet{Pols98}, applied by \citet{SOP97} and
now available in the public domain, were fitted to the positions of the stars
in the H--R diagram (see Fig.~\ref{fig:evol}). The selected tracks (assuming
solar abundances) fit the H--R positions of both stars quite decisively; they
suggest strongly that the secondary has already commenced its evolution away
from the main sequence, while the primary has apparently not yet commenced its
first travel up the red-giant branch.  The smallness of the error bars is
brought about by the high precision quoted for the {\it Gaia} parallax. The
values of the stellar masses given by the tracks are similarly tightly
constrained, and their ratio (1.318) reflects closely the value of $q$ (1.316)
determined by the SB2 solution (Table~\ref{tab:SB2}). Comparing these values
for the masses with those given in Table~\ref{tab:SB2} indicates that the
inclination of the orbit is about 32$^{\circ}$.

We then investigated the age of the component stars by finding the best-fitting
isochrone from the library made available in the public domain by
\citet{Pols98} (see Fig.~\ref{fig:iso}). The isochrone for log\,(age) of 8.60
fitted well to the H--R positions of both stars. We note that the age of this
binary is not a great as has been found for many other composite-spectrum
binaries analysed in this series of papers.

\begin{figure}[!h]
\centering
\includegraphics[height=2.2cm]{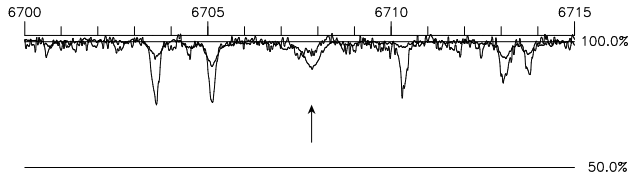}
\caption{The presence of Lithium $\lambda$\,6707\,\AA~in the spectrum of
HD~69479/80 (thick line), at the position of the arrow.  The comparison
spectrum (thin line) is that of o~Leo.}
\label{fig:lithium}
\end{figure}

If we now examine what this analysis shows regarding the evolutionary status of
the giant primary component, and in particular how `normal' it is, the
following facts become relevant: \\ (a) The value of log\,($L_1$) given in
Table~\ref{tab:parms} is close to the value of $\sim$1.7 listed by \citet{SK82}
for a G8\,III star but a little raised, as might be expected for a star that
had only recently arrived at the foot of the red-giant branch. \\ (b) The
system appears to be somewhat younger than is usually found for red giants that
are more advanced along the RG branch. \\ (c) The system has a short enough
period that the orbit might be expected to have reached full circularity (i.e.,
zero eccentricity). Instead, while the value of $e$ (Table~\ref{tab:SB2}) is
small it is probably significantly non-zero. \\ (d) Strong evidence of youth
is provided by the presence of a small but distinct feature at the position of
the $\lambda$\,6707\,\AA~Li~{\sc i} line.  At those red wavelengths the
observed spectrum of the binary is effectively that of the giant component
alone.  In order to confirm the identity of the Li feature, we compare in
Fig.~\ref{fig:lithium} the spectrum of HD~69479/80 with that of o~Leo, which
was found by \citet{G02} to contain two metallic-line stars; again, only the
cooler of the pair, having a spectral type of late Fm, is bright enough at
those wavelengths to record a signal.  It is uncommon to find Li {\sc i} strong
enough to be detected unambiguously in a class III giant later than mid-G, its
presence mostly being restricted to supergiants and bright giants. Its
appearance here definitely suggests that the giant star in this binary has
evolved across the Herzsprung gap but has not yet fully undergone the
upheavals, triggered by the onset of surface convection, that effectively
dilute lithium beyond detection.

From the above we conclude that the giant component is only about to commence
its first ascent of the red-giant branch, and is therefore well prior to
commencing the more advanced phase of He-burning and convective dilution in its
atmosphere. It is unusual that such a firm conclusion about the evolutionary
status of a giant star can be deduced so unambiguously, and it stems very
considerably from the high precision of {\it Gaia} data.

\section{The Binary System HD~69479/80}

Our analysis of HD~69479/80 has derived physical parameters for both component
stars.  Matching the H--R positions of the stars with standard evolutionary
tracks has confirmed that the secondary has already left the ZAMS, as was
suggested by its value of log\,$g$ = 3.75 (Fig.~\ref{fig:sub} and Section~3.1).
The cool-giant primary is currently positioned near the low-luminosity portion
of the red-giant branch, and the spectrum that best matched it (as judged from
the relatively clean residues that resulted from the subtraction process
described in Section~3.1) was that of the G8\,III standard 15~Cyg; accordingly
that was the classification that we gave it. Deriving a high-precision SB2
orbit, and fitting to evolutionary tracks and isochrones, yielded a mass of
2.9\,\msun~for the primary, 2.2\,\msun~for the secondary (a mean mass ratio $q$
of 1.317), and an age for the system of 8.60 Gyr. During its evolution the
system's orbit has become almost circular, and the secondary (an early-A dwarf)
has evolved away from the zero-age main sequence even though its mass is only
75\% of that of the cool giant.  The giant component shows a weak lithium
feature, which confirms our deduction that the star has not yet begun to ascend
the red-giant branch.

However, possibly the most remarkable aspect of this binary is its {\it lack}
of notable properties, rendering it highly important as a standard for analyses
of other composite-spectrum binaries. Although it has a period of only 91 days,
which \citet{Griffin11} demonstrated to be relatively short for a system
containing a cool giant (in our on-going study of composite-spectrum binaries,
only 20\% of the G--K giants have periods smaller than 100 days) the giant in
this binary shows no emission in the $H$ and $K$ or Balmer lines. Its rotation
rate -- determined by modelling the `dips' given by the {\it Coravel}
radial-velocity spectrometer -- is 6.5~\kms, which is as slow as for most
single late-G giants (though significantly faster than the value of
3.8~\kms~that would correspond to synchronous rotation). As a new addition to
the zoo of cool giants in binaries with relatively short periods, discussed at
some length by \citet{Griffin04}, this fact adds to the conclusion, already
apparent then, that the population of stars in these detached
composite-spectrum binaries is mixed, varied, and follows no set pattern or
restrictions that can be attributed exclusively to binarity. Thus, to have
found one whose components can be described as `normal' is a valuable asset.

\section{Acknowledgements}

We would like to acknowledge the support received from Mount Wilson Observatory
during the early days of this programme. REMG is grateful to the DAO (NRC) for
continuing privileges as a Visiting Worker, and for numerous observing runs
with the 1.2-m telescope, coud\'e spectrograph and CCD detector.  RFG is
pleased to acknowledge guest-investigator privileges at OHP, ESO, Palomar and
the DAO, and both authors thank the SERC and later PATT for defraying the
costs of observing runs.  We are pleased to acknowledge a helpful referee's
report, which led to several beneficial changes to the text of the paper, and
are grateful to Peter Stetson for discussions regarding stellar
parallaxes.

This research has made use of the SIMBAD data base, operated at CDS,
Strasbourg, France, and has also made use of data from the ESA mission
{\it Gaia} ({www.cosmos.esa.int/gaia}), processed by the {\it Gaia}
Data Processing and Analysis Consortium, DPAC. Funding for the DPAC
has been provided by national institutions, in particular the
institutions participating in the {\it Gaia} Multilateral Agreement.

\end{document}